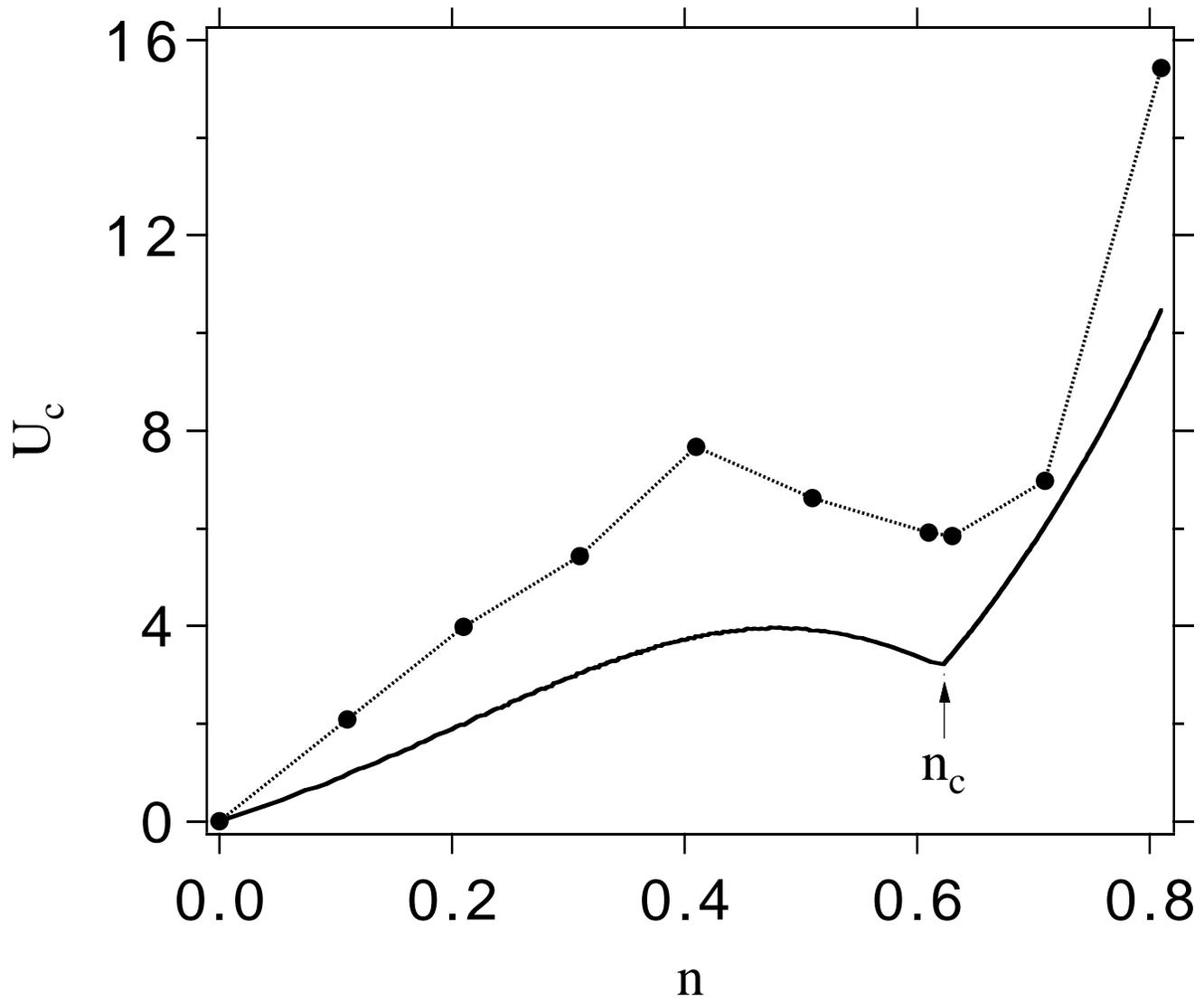

Fig. 1

# Variational study of ferromagnetism in the $t_1 - t_2$ Hubbard chain

S. Daul[a], P. Pieri[b], M. Dzierzawa[a], D. Baeriswyl[a], P. Fazekas[c]

[a] *Institut de Physique Théorique, Université, Pérolles, CH-1700 Fribourg, Switzerland.*

[b] *Dipartimento di Fisica, INFM and INFN, Università di Bologna, Via Irnerio 46, I-40126, Bologna, Italy.*

[c] *Research Institute for Solid State Physics, Budapest 114, POB 49, H-1525 Hungary.*

**Abstract**

A one-dimensional Hubbard model with nearest and (negative) next-nearest neighbour hopping is studied variationally. This allows to exclude saturated ferromagnetism for $U < U_c$. The variational boundary $U_c(n)$ has a minimum at a "critical density" $n_c$ and diverges for $n \to 1$.

The old question of the existence of ferromagnetism in the single-band Hubbard model has found a renewed interest, when rigorous proofs for a large-spin ground state were discovered for certain special cases [1, 2, 3, 4]. While it is in general very hard to show that the ground state is ferromagnetic, it is often rather easy to exclude saturated ferromagnetism, because the lowest-energy fully polarized $N$-electron state $|F\rangle$ is trivially constructed by occupying each of the $N$ lowest single-particle levels by a spin-up electron. To exclude saturated ferromagnetism one has then simply to find another state of lower energy. This approach has been applied to the Nagaoka problem [5] in terms of variational $N$-electron wave functions with a single overturned spin [6, 7, 8]. We have recently used such an approach to show that for non-pathological band structures and dimensions $d > 2$ the ground state is neither fully nor partially polarized in the low-density limit [9]. On the other hand a low-density route has been proposed for $d = 1$ by Müller-Hartmann [10]. It is this case that we want to discuss in the present contribution. We consider the $t_1 - t_2$ Hubbard chain described by the Hamiltonian

$$H = - \sum_{l=1,\sigma=\uparrow,\downarrow}^{L} (t_1 c_{l\sigma}^\dagger c_{l+1\sigma} + t_2 c_{l\sigma}^\dagger c_{l+2\sigma} + \text{h.c.}) + U \sum_{l=1}^{L} n_{l\uparrow} n_{l\downarrow} \qquad (1)$$

For $t_2 = 0$ we recover the simple Hubbard model with only nearest-neighbour hopping, for wich ferromagnetism has been discarded a long time ago by Lieb and Mattis [11]. For $t_2 \neq 0$ particles can pass each other and do not preserve a given order. The Lieb-Mattis theorem is then no longer applicable and indeed, for $U = \infty$ and $t_2 < 0$ ferromagnetism appears both for one hole in a half-filled band [12] and in the limit $t_2 \to 0$ for arbitrary densities ($n < 1$) [13]. These results raise the question about the range of parameters where the ground state is ferromagnetic. Before addressing this problem we turn to the easier question for which parameter values the fully polarized state is unstable.

The fully polarized state $|F\rangle$ is certainly unstable if the energy is reduced by flipping a single spin. We use two different trial states. The first ansatz [6], defined by

$$|\Psi\rangle = \prod_{l=1}^{L} (1 - \eta n_{l\uparrow} n_{l\downarrow}) c_{k_0\downarrow}^\dagger c_{k_F\uparrow} |F\rangle \qquad (2)$$

removes an up spin from the highest occupied level and puts a down spin to the minimum of the single-particle spectrum. It is important to note that this minimum moves away from $k = 0$ for large negative values of $t_2$, $t_2 < -\frac{|t_1|}{4}$, namely to $k_0 = \pm \arccos(\frac{|t_1|}{4})$. It is precisely in this region where according to Müller-Hartmann ferromagnetism should appear at low density and infinite $U$ [10]. The variational parameter $\eta$ in Eq. (2) reduces the double occupancy and steadily increases from 0 to 1 as a function of $U$. The energy change due to the spin flip consists essentially of a gain in band energy and a loss due to the Coulomb repulsion [9]. Thus one expects the fully polarized state to be unstable for small $U$. This is confirmed by explicit calculations shown in Fig. 1.

As a second trial state we use the wave function introduced by Edwards [14]

$$|\chi\rangle = \frac{1}{\sqrt{L}} \sum_{l=1}^{L} e^{iql} c_{l\downarrow}^\dagger \prod_{\alpha=1}^{N-1} c_{\alpha\uparrow}^\dagger(l)|0\rangle \tag{3}$$

where

$$c_{\alpha\uparrow}^\dagger(l) = \sum_{m=1}^{L} \varphi_\alpha(m-l) c_{m\uparrow}^\dagger \tag{4}$$

creates an up-spin electron in an orbital that is determined variationally. The variational parameters are the wave vector $q$ and the $(N-1)L$ amplitudes $\varphi_\alpha(l)$. The ansatz (3) is exact for $t_2 = 0$ [14]. For $t_2 \neq 0$ we add the Gutzwiller projection operator, which amounts to replace the one-particle orbitals $\varphi_\alpha(n)$ by $(1 - \eta \delta_{n,0})\varphi_\alpha(n)$. Then the wave function (3) includes the ansatz (2) as a special case. The variational energy for orthonormal one-particle orbitals reads

$$\begin{aligned} E(q, \{\varphi_\alpha(l)\}) &= 2t_1 \cos(q) \det S^{(1)} + 2t_2 \cos(2q) \det S^{(2)} + 2t_1 \text{tr} S^{(1)} \\ &\quad + 2t_2 \text{tr} S^{(2)} + U \sum_{\alpha\beta} \varphi_\alpha^*(0)\varphi_\beta(0) \end{aligned} \tag{5}$$

with overlap matrices $S_{\alpha\beta}^{(i)}$ defined as

$$S_{\alpha\beta}^{(i)} = \sum_{l=1}^{L} \varphi_\alpha^*(l) \varphi_\beta(l+i) \tag{6}$$

We have used the conjugate gradient method [15] to minimize the energy. The derivatives have been calculated analytically, and after each iteration the orbitals have been orthonormalized using the modified Gram-Schmidt method. As initial orbitals, which have to be close enough to the final wave functions $\varphi_\alpha(l)$, we use the $N-1$ lowest eigenfunctions for a single-site impurity. With this second ansatz, which involves much more variational freedom than the first, the instability line is moved to higher values of $U$, as shown in Fig. 1. It is worthwhile to add that a still more refined variational wave function [8] yields a similar instability line [16].

Our variational calculations delimitate a low $U$ region where the ground state is certainly not fully polarized. For the present model and large enough negative values of $t_2$ there is still a large region left where saturated ferromagnetism cannot be excluded, as illustrated in Fig. 1 for the particular value $t_2 = -0.8|t_1|$. In order to find out whether these variational lower bounds for the critical value $U_c$ are likely to be representative for the exact boundary, we have carried out exact diagonalizations for $L \leq 20$. The numerical results show that for densities around 0.4 and $t_2 \approx -0.5|t_1|$ the exact boundary lies typically about 20-50% above the variational bound of the second ansatz, Eq. (3). An interesting question is what happens above the "critical density" $n_c$ where the Fermi surface is reduced from four to two points. Both instability lines shown in Fig. 1 have a minimum at this density. In addition, for the Edwards ansatz the optimal wave vector $q$ is found to jump from $q = k_0$ (the location of the minimum in the band structure) for $n < n_c$ to $q = \pi$ for $n > n_c$. In our previous exact diagonalization studies (for an even number of particles and periodic boundary conditions) we tentatively associated $n_c$ with the phase boundary for ferromagnetism. This would indicate that the low-density route to ferromagnetism is not directly connected to the Nagaoka problem. In the mean-time we have extended our numerical studies both by imposing other boundary conditions and by increasing the chain length, using the density-matrix renormalization group method. The results obtained so far do not reproduce the phase boundary at $n_c$. Instead, at least for $U \to \infty$ and not too large $|t_2|$, ferromagnetism seems to exist for all densities $n < 1$. Further calculations will be needed to establish the full phase diagram.

Financial support by the Swiss National Foundation through the grant No. 20-40672.94 is gratefully acknowledged. We also thank E. Müller–Hartmann, K. Penc, K. Ueda and M. Sigrist for useful discussions and specially H. Castella for very helpful correspondence on technical details.

**Figure Captions**

Fig. 1. Variational phase diagram of the $t_1$–$t_2$ Hubbard chain in the $n$–$U$ plane, for $t_2 = -0.8|t_1|$. Full line : ansatz (2) for $L = 1000$, dots : ansatz (3) for $L = 100$.